



\font\bigbf=cmbx10 scaled\magstep2

\font\twelverm=cmr10 scaled 1200    \font\twelvei=cmmi10 scaled 1200
\font\twelvesy=cmsy10 scaled 1200   \font\twelveex=cmex10 scaled 1200
\font\twelvebf=cmbx10 scaled 1200   \font\twelvesl=cmsl10 scaled 1200
\font\twelvett=cmtt10 scaled 1200   \font\twelveit=cmti10 scaled 1200

\skewchar\twelvei='177   \skewchar\twelvesy='60


\def\twelvepoint{\normalbaselineskip=12.4pt
  \abovedisplayskip 12.4pt plus 3pt minus 9pt
  \belowdisplayskip 12.4pt plus 3pt minus 9pt
  \abovedisplayshortskip 0pt plus 3pt
  \belowdisplayshortskip 7.2pt plus 3pt minus 4pt
  \smallskipamount=3.6pt plus1.2pt minus1.2pt
  \medskipamount=7.2pt plus2.4pt minus2.4pt
  \bigskipamount=14.4pt plus4.8pt minus4.8pt
  \def\rm{\fam0\twelverm}          \def\it{\fam\itfam\twelveit}%
  \def\sl{\fam\slfam\twelvesl}     \def\bf{\fam\bffam\twelvebf}%
  \def\mit{\fam 1}                 \def\cal{\fam 2}%
  \def\tt{\twelvett}
  \textfont0=\twelverm   \scriptfont0=\tenrm   \scriptscriptfont0=\sevenrm
  \textfont1=\twelvei    \scriptfont1=\teni    \scriptscriptfont1=\seveni
  \textfont2=\twelvesy   \scriptfont2=\tensy   \scriptscriptfont2=\sevensy
  \textfont3=\twelveex   \scriptfont3=\twelveex 
 \scriptscriptfont3=\twelveex
  \textfont\itfam=\twelveit
  \textfont\slfam=\twelvesl
  \textfont\bffam=\twelvebf \scriptfont\bffam=\tenbf
  \scriptscriptfont\bffam=\sevenbf
  \normalbaselines\rm}



\def\beginlinemode{\endmode
  \begingroup\parskip=0pt 
\obeylines\def\\{\par}\def\endmode{\par\endgroup}}
\def\beginparmode{\endmode
  \begingroup \def\endmode{\par\endgroup}}
\let\endmode=\par
{\obeylines\gdef\
{}}
\def\singlespace{\baselineskip=\normalbaselineskip}
\def\oneandathirdspace{\baselineskip=\normalbaselineskip
  \multiply\baselineskip by 4 \divide\baselineskip by 3}

\def\doublespace{\baselineskip=
\normalbaselineskip \multiply\baselineskip by 2}

\newcount\firstpageno
\firstpageno=1
\footline={\ifnum\pageno<\firstpageno{\hfil}%
\else{\hfil\twelverm\folio\hfil}\fi}
\let\rawfootnote=\footnote              
\def\footnote#1#2{{\rm\singlespace\parindent=0pt\rawfootnote{#1}{#2}}}
\def\raggedcenter{\leftskip=4em plus 12em \rightskip=\leftskip
  \parindent=0pt \parfillskip=0pt \spaceskip=.3333em \xspaceskip=.5em
  \pretolerance=9999 \tolerance=9999
  \hyphenpenalty=9999 \exhyphenpenalty=9999 }
\def\dateline{\rightline{\ifcase\month\or
  January\or February\or March\or April\or May\or June\or
  July\or August\or September\or October\or November\or December\fi
  \space\number\year}}
\def\received{\vskip 3pt plus 0.2fill
 \centerline{\sl (Received\space\ifcase\month\or
  January\or February\or March\or April\or May\or June\or
  July\or August\or September\or October\or November\or December\fi
  \qquad, \number\year)}}


\hsize=6.5truein
\vsize=8.9truein
\voffset=0.0truein
\parskip=\medskipamount
\twelvepoint            
\oneandathirdspace           
\overfullrule=0pt       



\def\title                      
  {\null\vskip 3pt plus 0.2fill
   \beginlinemode \doublespace \raggedcenter \bigbf}

\def\author                     
  {\vskip 3pt plus 0.2fill \beginlinemode
   \singlespace \raggedcenter}

\def\affil                      
  {\vskip 4pt 
\beginlinemode
   \singlespace \raggedcenter \sl}

\def\abstract                   
  {\vskip 3pt plus 0.3fill \beginparmode
   \oneandathirdspace\narrower}

\def\endtitlepage               
  {\endpage                     
   \body}

\def\body                       
  {\beginparmode}               

\def\head#1{                    
  \vskip 0.25truein     
 {\immediate\write16{#1}
   \noindent{\bf{#1}}\par}
   \nobreak\vskip 0.125truein\nobreak}

\def\subhead#1{                 
  \vskip 0.25truein             
  \noindent{{\it {#1}} \par}
   \nobreak\vskip 0.15truein\nobreak}

\def\refto#1{[#1]}           

\def\references                 
  {\subhead{\bf References}         
   \beginparmode
   \frenchspacing \parindent=0pt \leftskip=1truecm
   \doublespace\parskip=8pt plus 3pt
 \everypar{\hangindent=\parindent}}

\gdef\refis#1{\indent\hbox to 0pt{\hss#1.~}}    

\gdef\journal#1, #2, #3, #4#5#6#7{               
    {\sl #1~}{\bf #2}, #3 (#4#5#6#7)}           

\def\refstylenp{                
  \gdef\refto##1{ [##1]}                                
  \gdef\refis##1{\indent\hbox to 0pt{\hss##1)~}}        
  \gdef\journal##1, ##2, ##3, ##4 {                     
     {\sl ##1~}{\bf ##2~}(##3) ##4 }}

\def\refstyleprnp{              
  \gdef\refto##1{ [##1]}                                
  \gdef\refis##1{\indent\hbox to 0pt{\hss##1)~}}        
  \gdef\journal##1, ##2, ##3, 1##4##5##6{               
    {\sl ##1~}{\bf ##2~}(1##4##5##6) ##3}}

\def\prd{\journal Phys. Rev. D, }

\def\cmp{\journal Commun. Math. Phys., }

\def\cqg{\journal Class. Quantum Grav., }

\def\endreferences{\body}

\def\figurecaptions             
  { \beginparmode
   \subhead{Figure Captions}
}

\def\endpage                    
  {\vfill\eject}

\def\endpaper                   
  {\endmode\vfill\supereject}

\def\endit
  {\endpaper\end}

\def\hook{\mathbin{\raise2.5pt\hbox{\hbox{{\vbox{\hrule height.4pt 
width6pt depth0pt}}}\vrule height3pt width.4pt depth0pt}\,}}
\def\today{\number\day\ \ifcase\month\or
     January\or February\or March\or April\or May\or June\or
     July\or August\or September\or October\or November\or
     December\space \fi\ \number\year}
\def\date{\noindent{\tt 
     Date typeset: \today\par\bigskip}}
\def\ref#1{Ref. #1}                     
\def\Ref#1{Ref. #1}                     

\def\frac#1#2{{\textstyle{#1 \over #2}}}
\def\half{{\textstyle{ 1\over 2}}}
\def\>{\rangle}
\def\<{\langle}
\def\eg{{\it e.g.,\ }}

\def\etc{{\it etc.}}

\def\sla{\raise.15ex\hbox{$/$}\kern-.57em}
\def\leaderfill{\leaders\hbox to 1em{\hss.\hss}\hfill}
\def\twiddle{\lower.9ex\rlap{$\kern-.1em\scriptstyle\sim$}}
\def\bigtwiddle{\lower1.ex\rlap{$\sim$}}
\def\gtwid{
\mathrel{\raise.3ex\hbox{$>$\kern-.75em\lower1ex\hbox{$\sim$}}}}
\def\ltwid{\mathrel{\raise.3ex\hbox
{$<$\kern-.75em\lower1ex\hbox{$\sim$}}}}
\def\square{\kern1pt\vbox{\hrule height 1.2pt\hbox
{\vrule width 1.2pt\hskip 3pt
   \vbox{\vskip 6pt}\hskip 3pt\vrule width 0.6pt}
\hrule height 0.6pt}\kern1pt}

\def\m@th{\mathsurround=0pt }
\def\leftrightarrowfill{$\m@th \mathord\leftarrow \mkern-6mu
 \cleaders\hbox{$\mkern-2mu \mathord- \mkern-2mu$}\hfill
 \mkern-6mu \mathord\rightarrow$}
\def\overleftrightarrow#1{\vbox{\ialign{##\crcr
     \leftrightarrowfill\crcr\noalign{\kern-1pt\nointerlineskip}
     $\hfil\displaystyle{#1}\hfil$\crcr}}}


\font\titlefont=cmr10 scaled\magstep3

\def\martinstyletitle                      
  {\null\vskip 3pt plus 0.2fill
   \beginlinemode \doublespace \raggedcenter \titlefont}

\font\twelvesc=cmcsc10 scaled 1200

\def\author                     
  {\vskip 3pt plus 0.2fill \beginlinemode
   \singlespace \raggedcenter\twelvesc}


\def\heading                            
  {\vskip 0.5truein plus 0.1truein      
\endheading
   \beginparmode \def\\{\par} \parskip=0pt \singlespace \raggedcenter}

\def\endheading
  {\par\nobreak\vskip 0.25truein\nobreak\beginparmode}

\def\subheading                         
  {\vskip 0.25truein plus 0.1truein     
   \beginlinemode \singlespace \parskip=0pt \def\\{\par}\raggedcenter}

\def\tag#1$${\eqno(#1)$$}

\def\align#1$${\eqalign{#1}$$}

\def\aligntag#1$${\gdef\tag##1\\{&(##1)\cr}\eqalignno{#1\\}$$
  \gdef\tag##1$${\eqno(##1)$$}}

\def\endaligntag{}

\def\overset #1\to#2{{\mathop{#2}\limits^{#1}}}
\def\underset#1\to#2{{\let\next=#1\mathpalette\undersetpalette#2}}
\def\undersetpalette#1#2{\vtop{\baselineskip0pt
\ialign{$\mathsurround=0pt #1\hfil##\hfil$\crcr#2\crcr\next\crcr}}}


\def\ref#1{Ref.~#1}                     
\def\Ref#1{Ref.~#1}                     
\def\[#1]{[\cite{#1}]}
\def\cite#1{{#1}}
\def\(#1){(\call{#1})}
\def\call#1{{#1}}
\def\taghead#1{}
\def\frac#1#2{{#1 \over #2}}
\def\half{{\frac 12}}

\def\12{{1\over2}}
\def\eg{{\it e.g.,\ }}

\def\etc{{\it etc.\ }}

\def\sla{\raise.15ex\hbox{$/$}\kern-.57em}
\def\leaderfill{\leaders\hbox to 1em{\hss.\hss}\hfill}
\def\twiddle{\lower.9ex\rlap{$\kern-.1em\scriptstyle\sim$}}
\def\bigtwiddle{\lower1.ex\rlap{$\sim$}}
\def\gtwid{\mathrel{\raise.3ex\hbox{$>$
\kern-.75em\lower1ex\hbox{$\sim$}}}}
\def\ltwid{\mathrel{\raise.3ex\hbox{$<$
\kern-.75em\lower1ex\hbox{$\sim$}}}}
\def\square{\kern1pt\vbox{\hrule height 1.2pt\hbox
{\vrule width 1.2pt\hskip 3pt
   \vbox{\vskip 6pt}\hskip 3pt\vrule width 0.6pt}
\hrule height 0.6pt}\kern1pt}
\def\tdot#1{\mathord{\mathop{#1}\limits^{\kern2pt\ldots}}}

\def\pmb#1{\setbox0=\hbox{#1}%
  \kern-.025em\copy0\kern-\wd0
  \kern  .05em\copy0\kern-\wd0
  \kern-.025em\raise.0433em\box0 }

\catcode`@=11
\newcount\tagnumber\tagnumber=0

\immediate\newwrite\eqnfile
\newif\if@qnfile\@qnfilefalse
\def\write@qn#1{}
\def\writenew@qn#1{}
\def\w@rnwrite#1{\write@qn{#1}\message{#1}}
\def\@rrwrite#1{\write@qn{#1}\errmessage{#1}}

\def\taghead#1{\gdef\t@ghead{#1}\global\tagnumber=0}
\def\t@ghead{}

\expandafter\def\csname @qnnum-3\endcsname
  {{\t@ghead\advance\tagnumber by -3\relax\number\tagnumber}}
\expandafter\def\csname @qnnum-2\endcsname
  {{\t@ghead\advance\tagnumber by -2\relax\number\tagnumber}}
\expandafter\def\csname @qnnum-1\endcsname
  {{\t@ghead\advance\tagnumber by -1\relax\number\tagnumber}}
\expandafter\def\csname @qnnum0\endcsname
  {\t@ghead\number\tagnumber}
\expandafter\def\csname @qnnum+1\endcsname
  {{\t@ghead\advance\tagnumber by 1\relax\number\tagnumber}}
\expandafter\def\csname @qnnum+2\endcsname
  {{\t@ghead\advance\tagnumber by 2\relax\number\tagnumber}}
\expandafter\def\csname @qnnum+3\endcsname
  {{\t@ghead\advance\tagnumber by 3\relax\number\tagnumber}}

\def\equationfile{%
  \@qnfiletrue\immediate\openout\eqnfile=\jobname.eqn%
  \def\write@qn##1{\if@qnfile\immediate\write\eqnfile{##1}\fi}
  \def\writenew@qn##1{\if@qnfile\immediate\write\eqnfile
    {\noexpand\tag{##1} = (\t@ghead\number\tagnumber)}\fi}
}

\def\callall#1{\xdef#1##1{#1{\noexpand\call{##1}}}}
\def\call#1{\each@rg\callr@nge{#1}}

\def\each@rg#1#2{{\let\thecsname=#1\expandafter\first@rg#2,\end,}}
\def\first@rg#1,{\thecsname{#1}\apply@rg}
\def\apply@rg#1,{\ifx\end#1\let\next=\relax%
\else,\thecsname{#1}\let\next=\apply@rg\fi\next}

\def\callr@nge#1{\calldor@nge#1-\end-}
\def\callr@ngeat#1\end-{#1}
\def\calldor@nge#1-#2-{\ifx\end#2\@qneatspace#1 %
  \else\calll@@p{#1}{#2}\callr@ngeat\fi}
\def\calll@@p#1#2{\ifnum#1>#2{\@rrwrite
{Equation range #1-#2\space is bad.}
\errhelp{If you call a series of equations by the notation M-N, then M and
N must be integers, and N must be greater than or equal to M.}}\else %
{\count0=#1\count1=
#2\advance\count1 by1\relax\expandafter\@qncall\the\count0,%
  \loop\advance\count0 by1\relax%
    \ifnum\count0<\count1,\expandafter\@qncall\the\count0,%
  \repeat}\fi}

\def\@qneatspace#1#2 {\@qncall#1#2,}
\def\@qncall#1,{\ifunc@lled{#1}{\def\next{#1}\ifx\next\empty\else
  \w@rnwrite{Equation number \noexpand\(>>#1<<) 
has not been defined yet.}
  >>#1<<\fi}\else\csname @qnnum#1\endcsname\fi}

\let\eqnono=\eqno
\def\eqno(#1){\tag#1}
\def\tag#1$${\eqnono(\displayt@g#1 )$$}

\def\aligntag#1\endaligntag
  $${\gdef\tag##1\\{&(##1 )\cr}\eqalignno{#1\\}$$
  \gdef\tag##1$${\eqnono(\displayt@g##1 )$$}}

\def\eqalignno#1{\displ@y \tabskip\centering
  \halign to\displaywidth{\hfil$\displaystyle{##}$\tabskip\z@skip
    &$\displaystyle{{}##}$\hfil\tabskip\centering
    &\llap{$\displayt@gpar##$}\tabskip\z@skip\crcr
    #1\crcr}}

\def\displayt@gpar(#1){(\displayt@g#1 )}

\def\displayt@g#1 {\rm\ifunc@lled{#1}\global\advance\tagnumber by1
        {\def\next{#1}\ifx\next\empty\else\expandafter
        \xdef\csname
 @qnnum#1\endcsname{\t@ghead\number\tagnumber}\fi}%
  \writenew@qn{#1}\t@ghead\number\tagnumber\else
        {\edef\next{\t@ghead\number\tagnumber}%
        \expandafter\ifx\csname @qnnum#1\endcsname\next\else
        \w@rnwrite{Equation \noexpand\tag{#1} is 
a duplicate number.}\fi}%
  \csname @qnnum#1\endcsname\fi}

\def\ifunc@lled#1{\expandafter\ifx\csname @qnnum#1\endcsname\relax}

\let\@qnend=\end\gdef\end{\if@qnfile
\immediate\write16{Equation numbers 
written on []\jobname.EQN.}\fi\@qnend}

\catcode`@=12

\catcode`@=11
\newcount\r@fcount \r@fcount=0
\newcount\r@fcurr
\immediate\newwrite\reffile
\newif\ifr@ffile\r@ffilefalse
\def\w@rnwrite#1{\ifr@ffile\immediate\write\reffile{#1}\fi\message{#1}}

\def\writer@f#1>>{}
\def\referencefile{
  \r@ffiletrue\immediate\openout\reffile=\jobname.ref%
  \def\writer@f##1>>{\ifr@ffile\immediate\write\reffile%
    {\noexpand\refis{##1} = \csname r@fnum##1\endcsname = %
     \expandafter\expandafter\expandafter\strip@t\expandafter%
     \meaning\csname r@ftext
\csname r@fnum##1\endcsname\endcsname}\fi}%
  \def\strip@t##1>>{}}

\def\citeall#1{\xdef#1##1{#1{\noexpand\cite{##1}}}}
\def\cite#1{\each@rg\citer@nge{#1}}	

\def\each@rg#1#2{{\let\thecsname=#1\expandafter\first@rg#2,\end,}}
\def\first@rg#1,{\thecsname{#1}\apply@rg}	
\def\apply@rg#1,{\ifx\end#1\let\next=\relax
\else,\thecsname{#1}\let\next=\apply@rg\fi\next}

\def\citer@nge#1{\citedor@nge#1-\end-}	
\def\citer@ngeat#1\end-{#1}
\def\citedor@nge#1-#2-{\ifx\end#2\r@featspace#1 
  \else\citel@@p{#1}{#2}\citer@ngeat\fi}	
\def\citel@@p#1#2{\ifnum#1>#2{\errmessage{Reference range #1-
#2\space is bad.}%
    \errhelp{If you cite a series of references by the notation M-N, then M 
and
    N must be integers, and N must be greater than or equal to M.}}\else%
 {\count0=#1\count1=#2\advance\count1 
by1\relax\expandafter\r@fcite\the\count0,
  \loop\advance\count0 by1\relax
    \ifnum\count0<\count1,\expandafter\r@fcite\the\count0,%
  \repeat}\fi}

\def\r@featspace#1#2 {\r@fcite#1#2,}	
\def\r@fcite#1,{\ifuncit@d{#1}
    \newr@f{#1}%
    \expandafter\gdef\csname r@ftext\number\r@fcount\endcsname%
                     {\message{Reference #1 to be supplied.}%
                      \writer@f#1>>#1 to be supplied.\par}%
 \fi%
 \csname r@fnum#1\endcsname}
\def\ifuncit@d#1{\expandafter\ifx\csname r@fnum#1\endcsname\relax}%
\def\newr@f#1{\global\advance\r@fcount by1%
    \expandafter\xdef\csname r@fnum#1\endcsname{\number\r@fcount}}

\let\r@fis=\refis			
\def\refis#1#2#3\par{\ifuncit@d{#1}
   \newr@f{#1}%
   \w@rnwrite{Reference #1=\number\r@fcount\space is not cited up to
 now.}\fi%
  \expandafter
\gdef\csname r@ftext\csname r@fnum#1\endcsname\endcsname%
  {\writer@f#1>>#2#3\par}}

\def\ignoreuncited{
   \def\refis##1##2##3\par{\ifuncit@d{##1}%
    \else\expandafter\gdef
\csname r@ftext\csname r@fnum##1\endcsname\endcsname%
     {\writer@f##1>>##2##3\par}\fi}}

\def\r@ferr{\endreferences\errmessage{I was expecting to see
\noexpand\endreferences before now;  I have inserted it here.}}
\let\r@ferences=\references
\def\references{\r@ferences\def\endmode{\r@ferr\par\endgroup}}

\let\endr@ferences=\endreferences
\def\endreferences{\r@fcurr=0
  {\loop\ifnum\r@fcurr<\r@fcount
    \advance\r@fcurr by 
1\relax\expandafter\r@fis\expandafter{\number\r@fcurr}%
    \csname r@ftext\number\r@fcurr\endcsname%
  \repeat}\gdef\r@ferr{}\endr@ferences}


\let\r@fend=\endpaper\gdef\endpaper{\ifr@ffile
\immediate\write16{Cross References written on 
[]\jobname.REF.}\fi\r@fend}

\catcode`@=12

\citeall\refto		
\citeall\ref		%
\citeall\Ref		%

\ignoreuncited
\def\proof{\bigskip\noindent{\bf Proof:\par}}
\pageno=0
\doublespace

\line{\hfill May 2003}
\title

The Helically-Reduced Wave Equation as a Symmetric-Positive System

\author
C. G. Torre
\affil
Department of Physics
Utah State University
Logan, UT 84322-4415 USA

\abstract
\doublespace
Motivated by the partial differential equations of mixed type that arise in the reduction of the Einstein equations by a helical Killing vector field, we consider a boundary value problem for the helically-reduced wave equation with an arbitrary source in 2+1 dimensional Minkowski spacetime.  The reduced equation is a second-order partial differential equation which is elliptic inside a disk and hyperbolic outside the disk. We show that the reduced equation can be cast into  symmetric-positive form.  Using results from the theory of symmetric-positive differential equations, we show that this form of the helically-reduced wave equation admits unique, strong solutions for a class of boundary conditions which include Sommerfeld conditions at the outer boundary. 
\endtitlepage

\body





\taghead{1.}
\head{1. Introduction}

Physical systems are typically governed by partial differential equations (PDEs) of a fixed type: elliptic, hyperbolic, or parabolic.  The mathematical properties of such equations have been extensively investigated (see, \eg Refs. \refto{Courant1962, Taylor1996}). Considerably less is known about PDEs of {\it mixed type}, by which we mean equations whose type is different in different subdomains of the domain of interest, \eg elliptic in one region and hyperbolic in another \refto{foot1}.  Compared to elliptic, hyperbolic or parabolic equations, mixed type equations are rather unusual,  both in the boundary conditions that can be imposed to get existence and uniqueness of solutions as well as in the regularity of solutions that are obtained. Moreover, the lower-order terms in equations of mixed type take on a more significant role than in equations of fixed type.  This latter feature means that it is difficult to obtain general results about PDEs of mixed type; to a large extent, one must investigate each set of equations, each set of boundary conditions, \etc separately.

In relativistic field theory on a fixed spacetime, mixed type equations occur after performing a symmetry reduction of hyperbolic PDEs with respect to an isometry group which has an infinitesimal generator that changes type from timelike to spacelike.  In generally covariant theories such symmetry reductions may yield PDEs of mixed type in appropriate gauges. An important example of the latter type, currently of considerable interest in gravitational physics, arises in the quasi-stationary approximation to the 2-body problem in general relativity \refto{Detweiler1992, Price2002, Friedman2002}. There one is interested in solving the Einstein equations for spacetimes admitting a helical Killing vector field.  The helical Killing vector field, which represents a rotating reference frame,  will be timelike near the bodies, and spacelike far from the bodies. The reduced Einstein equations (modulo gauge) can be expected to include non-linear PDEs of mixed type on the 2+1 dimensional manifold of orbits of the Killing vector field.  The reduced equations must be solved numerically, but one naturally desires as much {\it a priori} information about existence and uniqueness of solutions, regularity of solutions, admissible boundary conditions, \etc as one can get.  The quasi-stationary approximation to the relativistic 2-body problem is the principal motivation behind the work presented here.
 
As a warm-up for numerically solving the Einstein equations in the quasi-stationary approximation, the  wave equation and some of its non-linear extensions, reduced by the assumption of helical symmetry,  have been examined both analytically and numerically in \refto{Price2000}.  The helically-reduced wave operator is elliptic inside a ``light cylinder'' and hyperbolic outside the cylinder. It was shown in \refto{Price2000} that one can give a formal series solution of the helically-reduced wave equation for a source consisting of a pair of equal and opposite point charges, placed symmetrically with respect to the axis of helical symmetry. These analytical solutions satisfy Sommerfeld conditions at an outer boundary (which may be at infinity) \refto{foot2}.   The choice of such boundary conditions can be motivated on physical grounds, and the apparent analytic existence of unique solutions in the linear case and numerical solutions in the non-linear case gives confidence that the helically-reduced equation can be treated as a boundary value problem.  However, from a mathematical point of view it is not immediately clear {\it a priori} why such boundary conditions are admissible, that is, why one should expect unique solutions to exist.  Our goal here is to understand existence, uniqueness, regularity, \etc --- in short, the well-posed nature of this problem --- from a  general point of view that does not rely upon explicitly constructing a solution to the PDE. The idea is that such a point of view can be used to better understand the helically-reduced Einstein equations, which will not yield so easily to a direct assault.
 
In the past, certain boundary value problems of mixed type have been addressed using the theory of ``symmetric-positive'' differential equations \refto{Friedrichs1958}, which  can be viewed as a generalization of  elliptic and hyperbolic equations.  Friedrichs  \refto{Friedrichs1958} and Lax and Phillips \refto{Lax1960} have given the basic existence and uniqueness results for linear symmetric-positive equations.   In this note we shall show that the helically-reduced wave equation with arbitrary sources in 2+1 dimensions can be cast into symmetric-positive form. We can then deduce existence and uniqueness results for a class of boundary conditions that include the Sommerfeld conditions used in \refto{Price2000}.   These results provide support for the proposition that boundary value problems of mixed type arising from helical symmetry reductions --- such as arise in the relativistic two-body problem ---  are well-posed.

\taghead{2.}
\head{2. The helically-reduced wave equation}

We consider 2+1 dimensional Minkowski spacetime $({\bf R}^3,\eta)$ and a helical Killing vector field $\xi$. There will exist  an inertial-Cartesian coordinate chart $(t,x,y)$, 
such that
$$
\eta = -dt\otimes dt + dx\otimes dx + dy\otimes dy,
\tag
$$
and
$$
\xi = {\partial\over\partial t} + \Omega(x{\partial\over\partial y} - y{\partial\over\partial x})
\tag
$$
for some constant $\Omega$.
In the corresponding inertial-polar coordinates $(t,r,\phi)$ we have
$$
\eta = -dt\otimes dt + dr\otimes dr + r^2 d\phi\otimes d\phi,
\tag
$$
and
$$
 \xi = {\partial\over\partial t} + \Omega{\partial\over\partial\phi}.
\tag
$$
Note that $\xi$ is not of a fixed type:
$$
\eta(\xi,\xi) = r^{2}\Omega^{2}  - 1 \quad \cases{>0&\ for\  $r>{1\over\Omega}$\cr
=0&\  for\ $r={1\over\Omega}$\cr
<0&\ for\  $r<{1\over\Omega}$.}
\tag xinorm
$$
We will call the set of spacetime events with $r={1\over\Omega}$ the {\it light cylinder}.

The wave equation with source $j\colon {\bf R}^3\to {\bf R}$ is given by
$$
\square\Phi = j.
\tag wave_eq
$$
In the inertial-polar chart, the wave operator acting on a function $\Phi\colon {\bf R}^3\to{\bf R}$ takes the form
$$
\square\Phi = -\partial_t^2\Phi + {1\over r}\partial_r(r\partial_r\Phi) + {1\over r^2}\partial_\phi^2\Phi.
\tag
$$

We now restrict attention to fields and sources which are invariant under the 1-parameter isometry group $G$ generated by $\xi$. This is equivalent to requiring
$$
\xi(\Phi) = 0 = \xi(j).
\tag inv
$$
These conditions imply that $\Phi$ and $j$ define functions on the manifold of orbits ${\bf R}^3/G$, which shall be denoted by $\Psi$ and $f$, respectively.
Because the source $j$ is assumed $G$-invariant, and because $G$ is an isometry group for the spacetime, the wave equation \(wave_eq) admits $G$ as a symmetry group and defines a PDE relating $\Psi$ to $f$ on ${\bf R}^3/G$ \refto{Olver1993, AFT1999a}. To obtain this differential equation in local coordinates we proceed as follows. Group invariants on ${\bf R}^{3}$ are functions of $r$ and
$$
\varphi := \phi - \Omega t,
\tag varphi
$$
which define polar coordinates on ${\bf R}^3/G\approx {\bf R}^2$. In particular, granted \(inv), we have
$$
\Phi(t,r,\phi) = \Psi(r,\varphi),\quad j(t,r,\phi) = f(r,\varphi).
\tag quo
$$
The reduced field equation on ${\bf R}^3/G$ can be obtained by  substituting \(quo) into \(wave_eq),  which gives
$$
{1\over r}\partial_r(r\partial_r\Psi) + {1\over r^{2}}\chi(r) \partial_\varphi^2\Psi = f,
\tag psieq
$$
where
$$
\chi(r) = 1-\Omega^{2}r^{2}.
\tag chi
$$

Note that the light cylinder on ${\bf R}^3$ projects to a {\it light circle} at $r={1\over \Omega}$ on ${\bf R}^3/G$. Evidently, \(psieq) is elliptic inside the light circle and hyperbolic outside the light circle, which is a consequence of the changing character \(xinorm) of the Killing vector $\xi$.  Thus \(psieq) is a PDE of mixed type.  

In \refto{Price2000} \(psieq) is solved on a disk of radius $R$ with source $f$ corresponding to two equal and opposite ``scalar point charges'' placed symmetrically relative to the origin.  Sommerfeld conditions are imposed at the boundary of the disk;  and it is required that $\Psi$ vanishes at the origin. The solution is given as a formal infinite series. As noted in  \refto{Price2000}, despite the mixed type of the PDE and, in particular, despite the fact that a Sommerfeld condition was enforced at an outer boundary in the region where the PDE is hyperbolic, a unique solution exists. 
Here we provide a somewhat more general version of this result  using the theory of symmetric positive equations. We consider \(psieq)  on a domain \refto{foot3} $\epsilon\leq r\leq R$, $0<\epsilon<{1\over \Omega}$. We allow for a general source $f$ and we employ a class of boundary conditions that include Sommerfeld conditions  at $r=R$, such as  considered in  \refto{Price2000}. Specifically, we will impose the following boundary conditions:
$$
\Psi(\epsilon,\varphi) = 0,\quad \tau(\varphi) R\,\partial_r\Psi(R,\varphi) + \sigma(\varphi) \partial_\varphi\Psi(R,\varphi) = 0, \quad \sigma\tau\neq0
\tag psibc
$$
where $\sigma$ and $\tau$ represent smooth functions on the outer boundary $r=R$.
Sommerfeld boundary conditions correspond to setting $\tau=1/R$ and $\sigma = \pm \Omega$. 

Although we explicitly consider homogeneous boundary conditions \(psibc), because we allow for an arbitrary source $f$ in $\(psieq)$ a large class of inhomogeneous boundary conditions 
$$
\Psi(\epsilon,\varphi) = k(\varphi),\quad \tau(\varphi)\, R\,\partial_r\Psi(R,\varphi) + \sigma(\varphi)\, \partial_\varphi\Psi(R,\varphi) = l(\varphi), 
\tag inhom
$$
can also be accommodated. This is done by choosing a smooth function $\Lambda=\Lambda(r,\varphi)$ which satisfies the inhomogeneous boundary conditions \(inhom) and then redefining
$$
\Psi \to \tilde\Psi=\Psi - \Lambda,\quad f \to \tilde f=f - {1\over r}\partial_r(r\partial_r\Lambda) + {1\over r^{2}}\chi(r) \partial_\varphi^2\Lambda.
\tag 
$$
$\tilde\Psi$  now satisfies \(psieq) with source $\tilde f$ and homogeneous boundary conditions \(psibc), to which our results apply.

\taghead{3.}
\head{3. Symmetric-positive PDEs}

Existence and uniqueness results can be obtained for linear PDEs of mixed type if they can be cast into first-order, symmetric-positive form  with appropriate boundary conditions \refto{Friedrichs1958, Lax1960}.  Here we summarize the results from \refto{Friedrichs1958, Lax1960} which we shall need. 

For our purposes, the data used to define a symmetric positive system of equations with admissible boundary conditions will be taken to be \refto{foot4}:

\item{(i)}  A smooth manifold $M$ with smooth boundary $\partial M$; we set $\bar M= M \cup \partial M$. 
\item{(ii)}  A  smooth scalar density of weight-$1$ on  $M$, denoted by $\omega$. 
\item{(iii)}  A finite-dimensional real vector space $V$ with scalar product $(\cdot,\cdot)$. 

Let $u\colon \bar M\to V$.  We consider a first-order system of differential equations for $u$ on $M$ of the form
$$
Lu\equiv A^a\nabla_a u + Bu = h,
\tag FO
$$
where $h\colon \bar M \to V$ and, at each $x\in M$, $A^a$ and $B$ are linear transformations:
$$
A^a(x)\colon T^*_xM\times V\to V,\quad  B(x)\colon V\to V.
\tag
$$  
 For simplicity we assume that $A^a$ and $B$ depend smoothly on $x\in  M$.  The differential operator $\nabla_a$  is  the exterior derivative on functions defined by any basis for $V$.

\proclaim Definition 3.1. 
The system \(FO) is {\bf symmetric-positive} if: (1)  $A^a$  defines a symmetric operator  (with respect to the scalar product on $V$): 
$$
(A^a(x)v_a)^* = A^a(x)v_a,\quad \forall v\in T^*_xM \quad\ {\rm and}\quad \forall\  x\in M,
\tag symmetric
$$
and (2) the linear operator $K(x)\colon V\to V$, defined by
$$
K=B - \half \nabla_a (\omega A^a),
\tag kdef
$$ 
has a positive-definite symmetric part:
$$
K(x) + K^*(x) >0,\quad \forall x\in M.
\tag positive
$$ 

\noindent
We remark that $\nabla_a$ in \(kdef) is defined by its unique torsion-free extension to vector densities of weight-1 on $M$ taking values in $V\otimes V^*$.

A class of boundary conditions on $u\colon \bar M\to V$ have been determined such that there exist unique solutions to symmetric-positive systems of PDEs. Following Friedrichs, we call these boundary conditions {\it admissible}. They are defined as follows.

\proclaim Definition 3.2.
{\bf Boundary conditions} on $u\colon \bar M\to V$ are the requirement that,  at each $x\in \partial M$, $u(x)$ takes values in a linear subspace $N(x)\subset V$, which varies smoothly with $x$. 

We fix an outwardly oriented normal 1-form,  $n_a$, to the boundary $\partial M$. This 1-form is uniquely determined up to multiplication by a smooth positive function on the boundary.  We define \refto{foot5}
$$
\beta = n_a A^a\Big|_{\partial M}.
\tag
$$

\proclaim Definition 3.3. Let $ u\colon \bar M\to V$ be subject to the system of equations \(FO). The boundary conditions $u(x)\in N(x)$ on $\partial M$ are {\bf admissible} if  $N(x)$ is a maximal subspace such that the quadratic form $u\to (u,\beta u)$ is non-negative at each $x\in \partial M$.

Note that the admissibility of a set of boundary conditions does not depend upon the specific choice of outward covariant normal to the boundary. In the sequel we will make use of the following convenient characterization of admissible boundary conditions, which is due to Friedrichs \refto{Friedrichs1958}. 

\proclaim Proposition 3.4.
Admissible boundary conditions, $u(x)\in N(x)$ on $\partial M$ are equivalent to the linear boundary conditions $\beta_2 u=0$ on $\partial M$,  where $\beta_2$ arises from  a decomposition
$$
\beta = \beta_1 +\beta_2
\tag
$$
such, that for all $x\in \partial M$, (i) every $v\in V$ can be decomposed via
$$
v = v_1 + v_2, \quad \beta_1 v_2 =  \beta_2 v_1=0,
\tag
$$
and (ii)
$$
\mu := \beta_1 - \beta_2
\tag
$$
has a non-negative symmetric part:
$$
\mu+ \mu^*\geq 0.
\tag
$$


We now summarize the existence and uniqueness results of \refto{Friedrichs1958, Lax1960} for symmetric-positive systems.    We say that a mapping $u\colon M\to V$ is in  $L^2(M,V)$ if 
$$
||u||^2 \equiv\int_M (u,u)\omega <\infty.
\tag l2mv
$$ 


\proclaim Definition 3.5. Let the mappings $u\colon M\to V$ and $h\colon M\to V$ be in $L^2(M,V)$; $u$ is a  {\bf strong solution} to \(FO) satisfying the boundary conditions $u(x)\in N(x)$  on $\partial M$ if there exists a sequence of functions $\{u_k\}\in C^\infty(\bar M,V)$, satisfying the boundary conditions $u_k(x)\in N(x)$ on $\partial M$, such that
$$
u_k\to u,\quad {\rm and}\quad Lu_k \to h
$$
in the $L^2$ (semi-)norm \(l2mv).

\noindent


\proclaim Theorem 3.6 (Friedrichs, Lax \& Phillips).
If \(FO) is symmetric-positive  then it admits a unique, strong solution satisfying admissible boundary conditions.

We remark that the theorems appearing in \refto{Friedrichs1958, Lax1960} use (piecewise) continuously differentiable functions $u_k$ to define strong solutions. However it is straightforward to check that the relevant results go through for $\{u_k\}\in C^\infty(\bar M,V)$, which we use here.

\taghead{4.}
\head{4. A symmetric-positive system for the helically-reduced wave equation}

Here we show that the helically-reduced wave equation can be expressed in symmetric-positive form.
We choose $ M$ to be an annulus, $ M=\{(r,\varphi)|\epsilon< r< R\}$, equipped with the metric
$$
g= dr\otimes dr + r^2 d\varphi\otimes d\varphi,
\tag
$$
and associated density:
$$
\omega = (\det g)^{\half} = r.
\tag
$$
We set $V={\bf R}^2$ and equip it with the standard scalar product
$$
(u,v) = u_1 v_1 + u_2 v_2.
\tag
$$

We consider the following first-order system
$$
\eqalignno{
{1\over r}\partial_ru_2 + {1\over r^2}\chi \partial_\varphi u_1 &=  f,&(eq1)\cr
{1\over r}\partial_r u_1 - {1\over r^2}\partial_\varphi u_2 &= 0.&(eq2)
}
$$
Setting
$$
u_1 = \partial_\varphi \Psi ,\quad u_2=r\partial_r\Psi,
\tag psitou
$$
all classical ($C^2$) solutions to \(psieq), \(psibc) are solutions to \(eq1), \(eq2).  The solution $u$ thus obtained satisfies the boundary conditions
$$
u_1(\epsilon,\varphi) = 0,\quad \sigma(\varphi)u_1(R,\varphi) + \tau(\varphi) u_2(R,\varphi) = 0,\quad \sigma\tau\neq 0.
\tag ubc
$$
Conversely, given a classical ($C^1$) solution to  \(eq1), \(eq2) satisfying boundary conditions \(ubc), the function $\Psi$ defined by
$$
\Psi(r,\varphi) = \int_\epsilon^r dr^\prime {1\over r^\prime}u_2(r^\prime,\varphi),
\tag Psidef
$$
satisfies \(psitou) and hence \(psieq), \(psibc). In this sense the equations \(psieq), \(psibc) are equivalent to \(eq1), \(eq2), \(ubc). We write the system \(eq1), \(eq2) as
$$
(\tilde A^a\nabla_a u - \tilde h) = 0,
\tag oldeq
$$
where
$$
\tilde A^r = {1\over r}\left(\matrix{0 &1\cr 1&0}\right),\quad \tilde A^\varphi = {1\over r^2}\left(\matrix{\chi &0\cr 0 &-1}\right),
\quad
\tilde h=\left(\matrix{ f\cr 0}\right),
\tag
$$
and
$$
u = \left(\matrix{u_1\cr u_2}\right).
\tag 
$$

Now consider the following first-order system,
$$
L(\tilde A^a\nabla_a u - \tilde h) = 0,
\tag eqrel
$$
where
$$
L=\left(\matrix{a &-c\chi\cr c &a}\right),
\tag Ldef
$$
and $a=a(r)$ and $c=c(r)$ are smooth functions on $\bar M$. We require that $a$ and $c$  satisfy
$$
a^2 + c^2\chi \neq 0.
\tag nondeg
$$
so that \(eqrel) is equivalent to \(oldeq). We write \(eqrel) as
$$
A^a\nabla_a u =h,
\tag ueq
$$
where
$$
A^r= {1\over r}\left(\matrix{-c \chi &a\cr a &c}\right),
\quad
A^\varphi= {1\over r^2}\left(\matrix{a\chi &c\chi \cr c\chi&-a}\right),
\tag
$$
and
$$
h= \left(\matrix{a  f\cr c  f}\right).
\tag
$$

We now show that the functions $a$ and $c$ can be chosen so that the system \(ueq) is symmetric-positive.  The linear transformations defined by $A^a$ are symmetric for any choice of $a$ and $c$. Using \(kdef) we find that
$$
K = K^* =\half\left(\matrix{\partial_r (c\chi) &-\partial_r a\cr
-\partial_r a &-\partial_r c}\right).
\tag
$$
Necessary and sufficient conditions for \(positive) are
$$
\partial_r c <0,\quad (\partial_r c)\partial_r (c\chi) <-(\partial_r a)^2.
\tag spineq
$$
A pair of smooth functions that satisfy \(spineq) and \(nondeg) for $0<\epsilon\leq r\leq R$ are of the form
$$
a = const.\quad c = -\alpha + e^{-\Omega^3 r^3}
\tag acdef
$$
provided the constants $|a|$ and  $\alpha>0$ are chosen large enough, as is easily verified from 
$$
\partial_r c = -3\Omega^3 r^2 e^{-\Omega^3 r^3},
\quad
\partial_r(c\chi) = \Omega^2 r\Big\{2\alpha - e^{-\Omega^3 r^3}\left(2+ 3\Omega r\chi\right)\Big\}.
\tag
$$
Thus, with these choices for $a$ and $c$,  the equations \(ueq) are symmetric-positive. We summarize the preceding discussion as follows.

\proclaim Proposition 4.1. 
The equations \(psieq), \(psibc) are equivalent to the symmetric-positive system \(ueq), \(ubc), \(acdef).

\taghead{5.}
\head{5. Admissible boundary conditions}

We now consider the boundary conditions \(ubc) for which we have the following result.

\proclaim Proposition 5.1. 
The constants $a$ and $\alpha$ in \(Ldef) and  \(acdef) can be chosen so that the boundary conditions \(ubc) for \(ueq) are admissible.

\proof

  For the normal 1-form to the boundary we use
$$
n = \cases{\phantom{\ }Rdr &at $r=R$\cr
-\epsilon dr &at $r=\epsilon$.}
\tag
$$ 
We then have
$$
\beta(R) = \left(\matrix{-c(R)\chi(R) &a\cr a &c(R)}\right),\quad \beta(\epsilon) = \left(\matrix{c(\epsilon)\chi(\epsilon) &-a\cr-a &-c(\epsilon)}\right).
\tag
$$
 We consider boundary conditions of the form
$$
 \sigma u_1 + \tau u_2 =0\quad {\rm on}\ \partial M, 
\tag theubc
$$
where
$$
\sigma = 1,\quad \tau = 0,\quad {\rm at}\ r=\epsilon,
\tag epsilonbc
$$
and
$$
\sigma = \sigma(\varphi)\neq 0,\quad \tau=\tau(\varphi)\neq0,\quad {\rm at}\ r=R.
\tag Rbc
$$
Following Proposition 3.4, we set 
$$
\beta_1 = \pm N\left(\matrix{-\tau^2 c\chi - \sigma\tau a &\sigma\tau c\chi+\sigma^2 a\cr
\tau^2 a - \sigma\tau c & -\sigma\tau a + \sigma^2 c}\right),\quad
\beta_2=\pm N\left(\matrix{-\sigma^2 c\chi + \sigma\tau a &-\sigma\tau c\chi+\tau^2 a\cr
\sigma^2 a + \sigma\tau c & \sigma\tau a + \tau^2 c}\right),
\tag beta12
$$
where
$$
N = {1\over \sigma^2 + \tau^2},
\tag
$$
and
$$
\beta = \beta_1 +\beta_2.
\tag betasplit
$$
In  \(beta12) the plus/minus sign is to be used at the outer/inner boundary. The boundary conditions \(theubc) are equivalent to
$$
\beta_2 u = 0,\quad {\rm on}\ \partial M.
\tag
$$

At the inner boundary we have
$$
\half(\mu+\mu^*) = \left(\matrix{ -c(\epsilon)\chi(\epsilon) &0\cr
0&-c(\epsilon) 
}\right).
\tag
$$
We choose $\alpha$ sufficiently large so that $c(\epsilon)<0$ and we assume the inner boundary is within the light circle so we have $\chi(\epsilon) > 0$. Therefore $(\mu+\mu^*)$ is non-negative at the inner boundary; the inner boundary condition is admissible.  
At the outer boundary we have
$$
\half(\mu +\mu^*) =    N\left(\matrix{ -2 \sigma\tau a + (\sigma^2 -\tau^2)c\chi &-\sigma\tau \Omega^2 r^2 c\cr
-\sigma\tau \Omega^2 r^2 c &-2\sigma\tau a+(\sigma^2-\tau^2)c}\right)_{\scriptscriptstyle r=R}.
\tag symmu
$$
Evidently, the outer boundary conditions are admissible in the case $\sigma\tau>0$ provided $a$ is chosen sufficiently negative \refto{foot7}.  If $\sigma\tau<0$ the boundary conditions are admissible provided $a$ is chosen sufficiently positive \refto{foot7}. Thus the outer boundary conditions are admissible provided $\sigma\tau\neq 0$ there.



We remark that the admissibility of the outer boundary conditions did not depend upon the location of the outer boundary. Therefore the outer boundary conditions can be imposed outside, inside, and even {\it on} the light circle.  We also note that the proof of Proposition 5.1 shows that at the outer boundary neither Dirichlet conditions ($\sigma=1,  \tau=0$) nor Neumann conditions ($\sigma=0, \tau=1$) for $\Psi$ are admissible.

\taghead {6.}
\head{6. Conclusions}

From Propositions 4.1, 5.1 and Theorem 3.6 we have the following basic existence and uniqueness  result for the helically-reduced wave equation in its symmetric positive form. 

\proclaim Theorem 6.1.
The system \(ueq) admits a unique, strong solution satisfying the boundary conditions \(ubc). 

While this theorem only establishes existence of distributional solutions to the first-order form of \(psieq), \(psibc), it does imply that classical solutions  to \(psieq), \(psibc) are unique.  Considerable additional analysis is needed to establish existence of $C^2$ solutions to \(psieq), \(psibc). However, using Theorem 6.1 the  following regularity properties of $\Psi$ can be immediately inferred. 

Let us define $\tilde H^1(M)$ as the completion in the $H^1$ norm,
 $$
||\Psi||^2_1 = \int_M \left(\Psi^2 + g^{ab}\nabla_a\Psi \nabla_b\Psi\right)\omega,
\tag h1
$$  
 of the space of smooth functions $\Psi$ 
 satisfying the boundary conditions \(psibc).

We can then deduce the following from Theorem 6.1.
 \proclaim Corollary 6.2. 
 Let $u$ be the strong solution to \(ueq), \(ubc). The function $\Psi$, defined by
 $$
 \Psi =  \int_\epsilon^r dr^\prime\, {1\over r^\prime} u_2(r^\prime)\equiv I(u),
 \tag Psiandu
 $$
 is in $\tilde H^1(M)$ with (distributional) derivatives given by
 $$
 u_1 = \partial_\varphi \Psi,\quad  u_2 = r\partial_r\Psi.
 \tag
 $$
 
 \proof
 $I$ is easily verified to be a bounded linear transformation from  the dense subspace $C^\infty(M,V)\subset L^2(M,V)$ to $L^2(M)$, so $I$ can be extended to all of $L^2(M,V)$ thus defining $\Psi\in L^2(M)$ via \(Psiandu). Let $u_n\colon M\to V$ be the sequence of smooth maps that converges to the strong solution $u$. Clearly, 
$$
I(u_{2n})\Big|_{r=\epsilon} =0,  
\tag
$$
so that the inner boundary condition is satisfied. We have
$$
 \partial_r I(u_{2n}) = {1\over r} u_{2n} \to {1\over r} u_2\in L^2(M,V),  
\tag
 $$
and it is straightforward to verify that
$$
\partial_\varphi I(u_2n) = I(\partial_\varphi u_{2n}) \to u_1\in L^2(M,V).  
\tag
$$
 Thus $\Psi$ and its first derivatives are in $L^2$ and satisfy
 $$
 u_1 = \partial_\varphi \Psi,\quad  u_2 = r\partial_r\Psi,  
\tag
 $$
 so that the outer boundary conditions are satisfied as well.

 \square
\medskip

 Physically, the source $f$ in \(psieq) cannot be known with perfect precision. Furthermore, one may only have an approximately correct source appearing in a numerical solution. It is therefore important to note that the solution to \(ueq), \(ubc) depends continuously upon the source $h$, so small changes/errors in the choice of $h$ lead to correspondingly small changes in the solution $u$ (or $\Psi$). (Here ``small'' is defined by the $L^2$ norm.) To see this, we define a linear mapping 
 ${\cal S}\colon L^2(M,V)\to L^2(M,V)$ that associates a solution $u={\cal S}(h)$ of \(ueq), \(ubc) to each source $h$. The existence of this mapping follows from Theorem 6.1.  The uniqueness of strong solutions to  \(ueq), \(ubc)  implies  that the mapping $\cal S$ is closed and hence continuous by the closed graph theorem (see \eg \refto{Reed1980}).   From Corollary 6.2 the mapping $I\colon L^2(M,V)\to \tilde H^1(M)$ is bounded -- hence continuous --  and we then have the following corollary.

\proclaim Corollary 4.3.  
$\Psi\in \tilde H^1(M)$ defined by \(Psidef) depends continuously upon the source $f\in L^2(M)$.

\bigskip\noindent
{\bf Acknowledgment}

\refis{AFT1999a}{I. Anderson, M. Fels, C. Torre, \cmp 212, 653, 2000.}

\refis{Olver1993}{P. Olver, {\it Applications of Lie Groups 
to Differential Equations}, (Springer-Verlag, New York 
1993).} 

\refis{Lax1960}{P. Lax and R. Phillips, \journal Commun. Pure Appl. Math., 13, 427, 1960.}

\refis{Friedrichs1958}{K. O. Friedrichs, \journal Commun. Pure Appl. Math., 11, 333, 1958.}

\refis{Price2000}{J. Whelan, W. Krivan and R. Price, \cqg 17, 4895, 2000.}

\refis{Friedman2002}{J. Friedman, K. Uryu, M. Shibata, \prd 6, 064035, 2002.}

\refis{Price2002}{J. Whelan, C. Beetle, W. Landry and R. Price, \cqg 19, 1285, 2002.}

\refis{Detweiler1992}{J. Blackburn and S. Detweiler, \prd 46, 2318, 1992.}

\refis{Courant1962}{R. Courant and D. Hilbert, {\it Methods of Mathematical Physics}, Vol. 2 (Interscience, 1962).}

\refis{Taylor1996}{M. Taylor, {\it Partial Differential Equations}, (Springer 1996).}

Thanks to Chris Beetle for influential discussions and for digging up reference \refto{Friedrichs1958}. This work was supported in part by NSF grant PHY-0070867 to Utah State University.

\refis{CB2003}{C. Beetle, private communication.}

\refis{foot1}{For a survey of results on differential equations of mixed type, see  \eg 
M. Smirnov, {\it Equations of Mixed Type} (American Mathematical Society, 1978); 
J. Stewart, \cqg 18, 4983, 2001.}

\refis{foot2}{It should be noted that one of the goals in \refto{Price2000, Price2002} is to replace Sommerfeld conditions with ``minimum energy radiation balanced'' boundary conditions, which are non-local. It would appear that, generally speaking, nothing is known about the use of such boundary conditions with symmetric-positive equations.}

\refis{foot3}{The restriction to $\epsilon>0$  is needed to cast the field equations \(psieq) into symmetric positive form.}

\refis{foot4}{More general formulations are possible. For example,  symmetric-positive equations can be defined for sections of a vector bundle. See \refto{Friedrichs1958}.}

\refis{foot5}{This definition can be made more explicit by extending $n_a$ in an arbitrary fashion into $M$, then defining $\beta(x)$ as the limit of the smooth family of linear transformations $n_a(x)A^a(x)$ as $x$ approaches $\partial M$. The result is independent of the extension of $n_a$ into $M$.} 


\refis{foot7}{Recall that the equations \(ueq) are symmetric-positive provided $|a|$ is sufficiently large.}

\refis{Reed1980}{M. Reed and B. Simon, {\sl Methods of Modern Mathematical Physics, Vol. I}, (Academic Press, New York, 1980).}

\references

\endreferences

\endit
\bye